\documentclass[useAMS,usenatbib]{mnras} 
\usepackage{graphicx}
\usepackage{amssymb, amsmath}
\usepackage{times}
\usepackage{color}
\usepackage{lscape}
\usepackage[section]{placeins}
\usepackage{breqn}

\usepackage{lineno}

\usepackage{arydshln}

\usepackage{url}

\newcommand{\gtsima}{$\; \buildrel > \over \sim \;$}
\newcommand{\ltsima}{$\; \buildrel < \over \sim \;$}
\newcommand{\simgt}{\lower.7ex\hbox{\gtsima}}
\newcommand{\simlt}{\lower.7ex\hbox{\ltsima}}

\begin{document}


\title[Relativistic effects on Galaxy Target Selection]{Relativistic Effects on Galaxy Redshift Samples due to Target Selection}

\author[Alam et al.] {
    Shadab Alam$^{1,2,3}$ \thanks{Email: salam@roe.ac.uk \newline
https://sites.google.com/site/salamcosmology}, Rupert A. C. Croft$^{1,2}$, Shirley Ho$^{1,2,4,5}$, Hongyu Zhu$^{1,2}$ 
\newauthor\,
  and Elena Giusarma$^{1,2,4,5}$\\
    $^{1}$ Department of Physics, Carnegie Mellon University, 5000 Forbes Ave., Pittsburgh, PA 15213 \\
    $^{2}$ McWilliams Center for Cosmology, Carnegie Mellon University, 5000 Forbes Ave., Pittsburgh, PA 15213 \\
    $^{3}$ Institute for Astronomy, University of Edinburgh, Royal Observatory, Blackford Hill, Edinburgh, EH9 3HJ , UK \\
    $^{4}$ Lawrence Berkeley National Laboratory (LBNL), Physics Division, Berkeley, CA 94720-8153, USA \\
   $^{5}$ Departments of Physics and Astronomy, University of California, Berkeley, CA 94720, USA \\
}
    
\date{\today}
\pagerange{\pageref{firstpage}--\pageref{lastpage}}   \pubyear{2015}
\maketitle
\label{firstpage}

\begin{abstract}
\label{sec:abstract}
In a galaxy redshift survey the objects to be targeted for 
spectra are selected from a photometrically observed sample. 
The observed magnitudes and colours of galaxies in this parent sample
will be affected by their peculiar velocities, through
relativistic Doppler and relativistic beaming effects.
In this paper we compute the resulting 
expected changes in galaxy photometry.
The magnitudes of the relativistic effects are
a function of redshift, stellar mass, galaxy velocity and 
velocity direction.  We focus on the CMASS sample from the Sloan Digital Sky Survey (SDSS), Baryon Oscillation Spectroscopic Survey (BOSS), which is selected on the basis of colour and magnitude.
We find that 0.10\% of the sample ($\sim 585$ galaxies) has been
scattered into the targeted region of colour-magnitude space by 
relativistic effects, and conversely  0.09\% of the 
sample ($\sim 532$ galaxies) has been scattered out.
Observational consequences of these effects include an asymmetry in
clustering statistics, which we explore in a companion paper. Here
we compute a set of  weights which can be used to 
remove the effect of  modulations  introduced into the  density field
inferred from a galaxy sample.  
We conclude by investigating the possible effects of these relativistic modulation on large scale clustering of the galaxy sample.
\end{abstract}


\section{Introduction}
\label{sec:introduction}

General Relativity \cite[GR;][]{Einstein1916} combined with the standard 
cosmological model ($\Lambda$CDM) provides the  most successful theory of our 
universe with the minimum of external assumptions. The $\Lambda$CDM model paints a simple picture of structure formation arising from density fluctuations growing under  
gravity \citep{Comer1994}. For most of the Universe's history, these 
perturbations obey linear perturbation  theory 
\citep{Mukhanov1992,Liddle1993, Durrer1994, Ma1994, Bruni1994, Kopeikin2001,Bernardeau2002, Lagos2016}. The density field predicted by these theories have very specific statistical properties with multiple unique features \citep{1970ApJ...162..815P,Eis2005, Bassett2010,Coil2013}. We can measure most of the physical quantities of the universe just by comparing  one, two, three and higher point statistics of the predicted matter density field. Galaxies provide us with a window on the underlying matter density field of the universe. In the limit of linear perturbations,  galaxies can be assumed to form at the high-density peaks of the  underlying matter density field and should have same clustering properties up to a multiplicative constant (galaxy bias) \citep{Bardeen1986,Cole1989}.  Therefore creating three-dimensional maps of galaxies and studying their clustering properties provides one of the most precise ways to measure physical properties of our universe. In this paper, we address one of the complications of making these
maps from galaxy redshift surveys which is usually ignored: the effect of
peculiar velocities on galaxy photometry and thus the target selection.

Carrying out large galaxy surveys has been a challenging task, which was
made easier by the development of CCD cameras \citep{1998ASSL..228.....B}. Many astronomy projects were involved in the development and adoption of CCD technology for telescopes \citep{Arnaud1994, Abe1997, Bauer1998, Boulade1998, Fukugita1996, Gunn1998}. These have led to various photometric surveys covering increasingly large parts of sky with improved depth and resolution \cite[][DES\footnote{\url{http://www.darkenergysurvey.org/survey/}}]{York2000, Gladders2005, Kaiser2010,Takada2010, Gilbank2011}. Such surveys provide an excellent map of the angular distribution of  galaxies,  but precise measurements of the cosmological line-of-sight distance, and hence creation of three dimensional maps, requires redshifts ($z$). The redshift quantifies the wavelength shift
of features in galaxy spectra and hence requires observing galaxy's spectral energy distributions (SED). The measurement of galaxy SED requires targeting each galaxy individually and is a very expensive process. An early large galaxy redshift surveys was the CfA redshift survey \citep{CfaSurvey1989} which observed 22000 galaxies one at a time. Galaxy surveys targeting much large numbers of galaxies for SED measurement became possible with the advent of optical fibres combined with the ability to observe hundreds of SEDs in a single exposure. The huge increase in the number of spectra that we could observe started the era of large galaxy redshift surveys e.g: \citet[ LCRS]{1996ApJ...470..172S},  \citet[2dF]{Colless2003},  \citet[6dF]{Jones2009}, \citet[SDSS-III]{Eisenstein2011},  \citet[WiggleZ]{WiggleZ},  \citet[DEEP2]{Deep2013},  \citet[VIPERS]{Garilli2014},   \citet[GAMA]{gama2015}.

To make this process efficient, it is important to have prior knowledge about the location of possible targets. Therefore, generally galaxy redshift surveys require samples of objects observed photometrically to serve as parent sample. Various algorithms and knowledge of galaxy evolution models are employed to create sub-samples of such parent samples to be targeted for spectra (for example \citet{Reid2016}). Generally, these selection algorithms use various magnitude and colour cuts to define these subsamples. We know that the observed magnitudes and colours of galaxies are affected by their peculiar motion \citep{Teerikorpi1997}. This can influence the final spectroscopic galaxy target sample obtained after following the target selection rules \citep{Kaiser2013}. Such effects will  act to modulate the observed galaxy density in the observed sample, in a way which will be correlated with galaxy properties including redshift, mass and velocity. This could in principle introduce new features into the measured clustering of galaxies and also bias the physical properties inferred from such clustering observations.

In this paper, we examine the special relativistic effects that galaxy
peculiar velocity have on their observed SEDs and the photometric
quantities derived from them. We then discuss the impact of these
effects on an observed sample of galaxies. We use the Sloan Digital
Sky Survey III (SDSSIII) Baryon Oscillation Spectroscopy Survey (BOSS)
CMASS sample from Data Release 12 (DR12) as an example to show how
relativistic effects will impact target selection which uses cuts in
the magnitude and colour plane.  We then discuss how these introduce
density modulation in the observed sample.  We define a weighting
scheme to compensate for such modulation and look at its effect on the
clustering signal. We conclude with a discussion about the impact of
such effects on the large scale structure analyses.  We note that we
restrict ourselves here to the effect of peculiar velocities on
spectroscopic target selection. This is distinct from the effect of
velocities on the properties of galaxies inferred from the
spectroscopic sample \cite[e.g.,][]{2015MNRAS.450..883K,
  2014MNRAS.443.1900B}. We would like to stress here
  that the main focus of and motivation for the paper is the derivation of the
  relativistic weights. The impact on the clustering signal is just one of
  the areas which can be assessed using these weights. We are
  most interested however in the weights themselves, which can be 
  used to  model the impact of
  relativistic effects (specifically relativistic beaming) 
 on galaxy clustering. We investigate this aspect in our companion paper
  \citet{Zhu2016Nbody}.

\section{Effects of peculiar velocities on galaxy spectra}
\label{sec:theory}
We  study the relativistic effects of galaxy motion on galaxy spectra and how  they affect observed galaxy flux and colour. This will help us estimate the impact of such observational effects on our final observed samples. We consider two kinds of effects. The first is the redshift or blueshift applied to the spectrum due to relative motion between the observer and galaxy. The second  is the change in flux coming from relativistic boost and beaming.  Note that we do not consider the impact of magnification caused by gravitational lensing \citep{2012ApJ...744L..22S, 2010MNRAS.405.1025M}.

\subsection{Relativistic Doppler effect}
The relativistic Doppler effect shifts the observed wavelength of a photon 
with respect to the emitted wavelength in a manner which depends on the line of sight velocity of the source. The observed wavelength and emitted wavelength for a galaxy moving along the line-of-sight are related by the following equation, where $\beta_{los}=v_{los}/c$ is the ratio of line of sight velocity ($v_{los}$) and the speed of light ($c$):
\begin{equation}
\lambda_o = \lambda_e \sqrt{\frac{1-\beta_{los}}{1+\beta_{los}}}.
\label{eq:lamshift}
\end{equation}
Here $\lambda_o$ and $\lambda_e$ are the observed and emitted wavelengths respectively. The galaxy's velocity along the line of sight consists of two components. First component is the Hubble velocity due to the  expansion of the universe (denoted by $v_e$) while the second component is due to local dynamics, the peculiar velocity and denoted by  $v_p$. The total line-of-sight velocity of a galaxy $v_{los}$ is given by relativistic addition of the two components under the assumption of negligible matter density so that
\begin{equation}
v_{los}= \frac{v_e + v_p}{1+ \frac{v_e v_p}{c^2}}.
\label{eq:addvel}
\end{equation}

The expansion of the Universe acts to redshift the
galaxy spectrum, and peculiar velocities lead to
additional shifts.
This implies that photometric bands see
different parts of the spectrum for galaxies with different
redshifts. Accounting for this shift leads to the well known
K-correction, \citep[see for example the case of massive
galaxies][]{Hogg2002, Blanton2003a}. 
In order to apply a K-correction to galaxy magnitudes in different bands,
it is necessary to use an estimate of the galaxy redshift. In the present
paper, we concern ourselves with target selection for galaxy spectroscopic
redshift 
surveys, and we assume that this target selection is carried out
using galaxy magnitudes before a redshift is known, and hence without
K-corrections. Photometric redshifts could instead be used to 
compute K-corrections first, but we consider surveys such 
as BOSS/CMASS \citep{Reid2016} and the SDSS main galaxy sample \citep{Strauss2002} where this is not done.

 First, we note that the effect of
shift in wavelengths due to different components of the galaxy
velocity can be separated as follows:

\begin{equation}
\left(\frac{\lambda_o}{\lambda_e}\right)^2 =\left( \frac{1-\beta_{los}^e}{1+\beta_{los}^e} \right) \left(\frac{1-\beta_{los}^p}{1+\beta_{los}^p} \right)
\label{eq:vel5}
\end{equation}
Equation \ref{eq:vel5} shows that the
Doppler shifts in wavelength due to different velocity components is
separable and hence justifies our treatment to separate peculiar
velocity from the Hubble velocity due to the expansion of the Universe.
We note that additional terms such as that due to 
the gravitational redshift/Sachs-Wolfe effect are also relevant, and
are treated in our companion paper \citep{Zhu2016Nbody}. To linear order
in perturbation theory, the combined effects are described in detail
by e.g., \cite{Yoo2014,Bonvin2014a}.
We also note that if galaxy band magnitudes were 
K-corrected using the 
observed galaxy redshift this would take into account the 
effect of peculiar velocities as well as the Hubble expansion and other 
components. As stated above, such K-corrections are not
relevant for the target selection considered here.
 
 It is important to define the sign convention for velocity to avoid any
confusion. From now on we use positive velocity and $\beta$ to
indicate that the line-of-sight component of galaxy peculiar velocity
is toward the observer. Negative velocity will imply that the galaxy's
line-of-sight component of velocity is moving away from the observer. 
In the situation when a galaxy is moving with velocity $c\beta$ at an angle $\theta$ from the
line-of-sight then the Doppler shift will have an additional term due to
the  transverse velocity. 
The observed wavelength and emitted wavelength for a galaxy moving in such a situation is given by the following equation, where $\gamma=1/\sqrt{1-\beta^2}$

\begin{equation}
\lambda_o =  \gamma (1-\beta \cos(\theta))\lambda_e
\label{eq:lamshift-all}
\end{equation}

\subsection{Relativistic Beaming effect}
Relativistic beaming modifies the apparent brightness of a galaxy due to its peculiar motion. The peculiar motion of galaxy through the Doppler shift modifies the energy of emitted photons and the number of photons emitted per unit time.  The direction in which photons are  emitted is also different in the observed frame compared to the galaxy's rest frame, leading to an anisotropic pattern of  emission in the observer's frame. Taken together, these effects  are known as relativistic beaming.  The effect on the spectral brightness  can be derived using special relativity. The spectral brightness ($I_\nu$) of a galaxy is defined to be the energy observed per unit time, per unit area of the detector, per unit frequency and per unit solid angle\footnote{More discussion in \citet{Hogg1997}. Section 7.4 of \url{http://cosmo.nyu.edu/hogg/sr/sr.pdf} is most relevant}:

\begin{equation}
I_\nu = \frac{\Gamma E}{\sigma \Omega},
\label{eq:brightness}
\end{equation}
where $\Gamma$ is the number of photons emitted per unit time, $E$ is the energy of emitted photons, $\Omega$ is the solid angle subtended by the observed galaxy and $\sigma$ is the area of the detector. Each of the quantities appearing in equation \ref{eq:brightness} will be modified by the peculiar motion of the galaxy in the observed frame. The spectral brightness in the observed (telescope) frame ($I_{\nu}^o$) and emitted (galaxy rest) frame ($I_{\nu ^\prime}^e$) are related by following equation:

\begin{equation}
\frac{I_{\nu_o}^o}{I_{\nu_e}^e} = \left(\frac{\nu_o}{\nu_e} \right)^3 =\left[\gamma (1-\beta cos(\theta)) \right]^{-3}.
\label{eq:Ioenu}
\end{equation}  

Here the Lorentz factor $\gamma=\frac{1}{\sqrt{1 - \beta^2}}$  and $\theta$ is the angle the velocity vector makes with the line of sight direction. The above expression is derived using the fact that phase space volume is invariant under Lorentz transformations. It is proportional to the number of photon in a quantum state. This makes the quantity  $\frac{I_{\nu}}{\nu^3}$ Lorentz invariant and leads to equation\footnote{A detailed derivation of these equations can be found in \citet{Goodman2013}. Chapter 1 of \url{http://www.astro.princeton.edu/~jeremy/heap.pdf} is most relevant} \ref{eq:Ioenu}.
This equation is in terms of flux per unit frequency whereas our measurements will be in flux per unit wavelength. The spectral brightness per unit  frequency ($I_\nu$) can be converted to the spectral brightness per unit wavelength ($I_\lambda$) using: 

\begin{equation}
I_{\lambda}=\frac{dF}{d \lambda} = \frac{dF}{d \nu} \frac{d \nu}{d \lambda}=\frac{I_{\nu}}{ \lambda^2}
\label{eq:lnu}
\end{equation} 
Where we have used $\nu \lambda=c$. 

Finally, the observed and emitted spectral brightness per unit wavelength can be obtained by combining equations \ref{eq:lamshift-all}, \ref{eq:Ioenu} and \ref{eq:lnu}:

\begin{equation}
\frac{I_{\lambda_o}^o}{I_{\lambda_e}^e} = \left[\gamma (1-\beta cos(\theta)) \right]^{-5}
\label{eq:Ioel}
\end{equation}

It is important to note that relativistic beaming depends on both the magnitude and direction  of  the source velocity and not just its the line-of-sight component. 

\subsection{Effects of velocity on the observed spectra}

\begin{figure}
\includegraphics[width=0.5\textwidth]{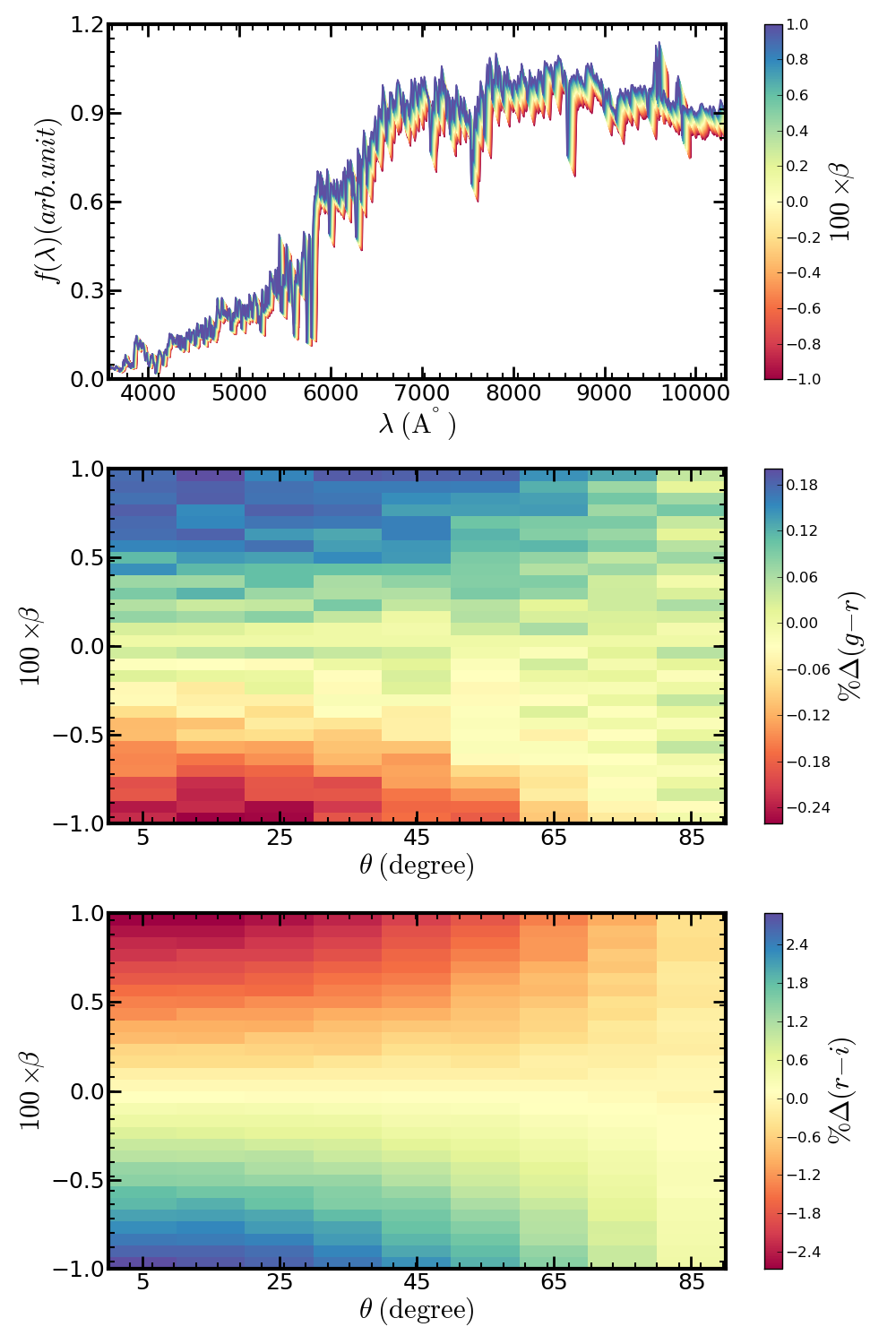}
\caption{The relativistic effects on the spectra and observed colour of a single galaxy. The top panel shows the flux of a galaxy SEDs on the y-axis, with x-axis showing wavelength in \r{A} and the colour scale showing  velocity. 
 Two effects are liiustrated, the first being the wavelength shift and the second being the rescaling of flux for the same wavelength as the source galaxy moves towards or away from the observer. The middle and bottom panel show the percentage change in the $g-r$ and $r-i$ colours as a function of the magnitude and direction of the galaxy velocity respectively. The change in colours are strongest when galaxy velocity is alight towards the line-of-sight ({\it i.e.} $\theta=0$), and vanishes when the galaxy velocity become perpendicular to the line-of-sight ({\it i.e.} $\theta=90$). }
\label{fig:OneSpectra}
\end{figure}

The spectra observed for a  galaxy redshift survey experience both the effects discussed in the previous two subsections: the shift in wavelength due to Doppler shift and  the change in flux due to relativistic beaming. 

To compute these effects on the broad band magnitudes used for
target selection we can make use of
some template galaxy spectra and redshift them. 
The spectra observed
from the BOSS/CMASS survey can fulfill this purpose. We therefore 
now describe how the BOSS/CMASS fibre spectra are affected by peculiar 
velocities.

The following equation describes how the observed flux per unit wavelength ($f_{\lambda}^o$) 
is related to the emitted flux per unit wavelength  ($f_{\lambda}^e$) at wavelength ($\lambda_e$),
as a function of observed wavelength ($\lambda_o$)

\begin{equation}
f_{\lambda}^o (\lambda_o,\beta,\theta) = f_{\lambda}^e (\lambda_e) \left[\gamma (1-\beta \cos(\theta)) \right]^{-5} 
\label{eq:spectra}
\end{equation}

Here the galaxy is moving with peculiar velocity $v=\beta c$ along the
direction at angle $\theta$ from the line-of-sight. The observed ($\lambda_o$) and
emitted ($\lambda_e$) wavelengths are related by equation \ref{eq:lamshift-all}.
While deriving equation \ref{eq:spectra}, we have assumed isotropic emission of light 
from galaxies. In the case of realistic galaxies the different components of galaxies  can have a non-isotropic emission pattern. 
In such cases the shape of the galaxy can be aligned with tidal forces acting on it (causing so-called intrinsic alignments) and hence may show a correlation with the peculiar velocity. Modelling the anisotropic emission from galaxies and its correlation with peculiar velocity is beyond the scope of this paper but could be studied in future work.

Figure \ref{fig:OneSpectra} shows the effect of relativistic beaming and relativistic doppler shift on the observed galaxy spectra and colour. The top panel focuses on the galaxy SED. The x-axis shows the wavelength in \r{A} and the y-axis shows the observed flux. The colour scale represents the velocity of the galaxy in the unit of speed of light. The spectrum corresponding to $\beta=0$ represents the emitted galaxy spectrum. We can clearly see the two effects discussed in the previous two sections. The relativistic Doppler shift causes the atomic lines to shift in wavelength. Relativistic beaming increases the observed flux for positive $\beta$ (moving towards the observer) and decreases it for negative $\beta$ (moving away from the observer). The middle and bottom panels show the percentage change in the $g-r$ and $r-i$ colour as a function of different velocity magnitude (varying along the y-axis) and  velocity direction with respect to the  line-of-sight (x-axis). The percentage change in $g-r$ colour is at the level of 0.2\% when the galaxy has a peculiar velocity of 3000 ${\rm km\,s}^{-1}$.
For realistic velocities of around 400 ${\rm km\,s}^{-1}$ (See section 4.5) the change is around 0.05\% . For  $r-i$ colour the percentage change is significantly higher,
at the level of 3\% for 3000 ${\rm km\,s}^{-1}$ galaxies and $\sim 0.5\%$ for 400 ${\rm km\,s}^{-1}$.
This difference between colour bands illustrates
that the strength of the relativistic selection effects 
will depend on galaxy spectrum and hence galaxy type in a relatively complex
way.

\section{Effects of velocities on Selected Catalog}
\label{sec:TS}

Most large
galaxy redshift surveys feature a two-step process of photometric
target selection and spectroscopic follow-up. Grism spectroscopy and other 
techniques for one-step generation of galaxy redshift samples
have been used in the past \cite[e.g. ][]{1996MNRAS.279.1057S, 2015arXiv151002106M,2008ASPC..399..115H} and will play a prominent role
in the future (EUCLID: \citet{2010SPIE.7731E..2YC}, WFIRST: \citet{2013arXiv1305.5425S},  SPHEREx: \citet{2016AAS...22714701B} ). 
Neverthless, fibre spectrographs are also becoming larger 
and photometric selection of galaxy targets will be used to generate   
samples of tens of millions of galaxy redshifts in the next few years 
\citep{2016arXiv161100036D}. We therefore focus in this paper on photometric
target selection.

In order to obtain a reasonable target sample one must determine the 
properties of each object based on photometric magnitudes. This require 
detailed modeling of the SEDs of different kind of objects. 
The targets of interest are then selected from a photometric sample which 
has predefined depth and redshift
 coverage.
Historically target selection was the result of
 simple magnitude cuts. Recent redshift surveys employ more complex
 sample selection with various cuts in the colour-magnitude plane \citep{Reid2016, 2016ApJS..224...34P}. The final observed samples will also be affected
by several biases due to the interplay between the sharp magnitude cut,
the luminosity function  and errors
in  the observed magnitudes. These biases are well understood and 
discussed in detail by e.g., \citet{Teerikorpi1997}. We are not focusing on 
biases of such kind, but instead we are concerned about the modulations 
introduced in the inferred density field due to galaxy peculiar motion, 
but distinct from redshift space distortions. As mentioned 
in Section 2,
we deal exclusively with target selection where
K-corrections have not been applied to galaxy broad-band magnitudes before
targets are selected. 
If redshifts are available and those corrections are made, the effects
of peculiar velocities on galaxy colours would be nullified by the 
K-correction.

\subsection{Magnitude limited sample}

A magnitude limited sample is  one which has been selected only by 
applying a limiting magnitude cut. The effect of peculiar velocities 
on such a 
sample is relatively simple to understand. The galaxies  moving towards the observer will have their magnitudes boosted and those that are intrinsically
just below the threshold will move into the sample. The galaxies moving away from an observer will have their magnitudes suppressed and  hence those just above the magnitude limit will move out of the sample. We can therefore construct a simple picture in which the probability of a galaxy passing the sample cut is determined its velocity. The constant of proportionality will depend on the true magnitude of the galaxy and its spectrum and it will always be positive. This means galaxies moving towards the observer will always have a higher probability of making the sample cut compared to galaxies moving away from the observer. This is true unless one considers an exotic galaxy SED, for example, an SED  in which flux decreases with wavelength fast enough, such that the gain in flux by relativistic beaming is smaller than the reduction in flux caused by relativistic Doppler effect.

\subsection{colour-Magnitude cuts}

Most of the current and future galaxy redshift survey have a more 
complicated targeting algorithm than simple magnitude cuts.
In a more complicated scenario where the sample selection has several colour and magnitude cuts, the simplest expectation that galaxies moving towards the observer will have a higher probability of making into the sample does not hold true. The exact nature of cuts, details of spectra and the galaxy population can lead to the probabilities of including galaxies moving towards the observer being  smaller than those moving away from the observer. Such effects depend on the redshift, halo mass and peculiar velocity (both magnitude and direction) of the observed galaxy. This can lead to extra structure in the number density of the observed target and affect the clustering measurements. This has been assumed to be unimportant for current and future surveys. 
We will investigate the validity of this past assumption. 
Some analyses of galaxy clustering
rely on partitioning a sample into subsamples based on their observed
properties \citep{2006MNRAS.369...68S, Croft2013, Alam2016Measurement}. The
effects we model in this paper are likely to be relatively more important
for these analyses, as they will have different strengths for
sub-samples with different galaxy properties.

\section{Special Case: SDSS III CMASS Sample}
\label{sec:CMASS}
The SDSS III CMASS sample is one of the key target datasets where we have a large number of massive galaxies with photometric and spectroscopic observations. We use this sample as an example, computing the effects of relativistic beaming
and Doppler shifting in detail. This analysis can be easily extended to other surveys. We first briefly describe the sample and introduce the relevant quantity necessary to understand the CMASS target selection.

\subsection{CMASS Sample}
\begin{figure}
\includegraphics[width=0.5\textwidth]{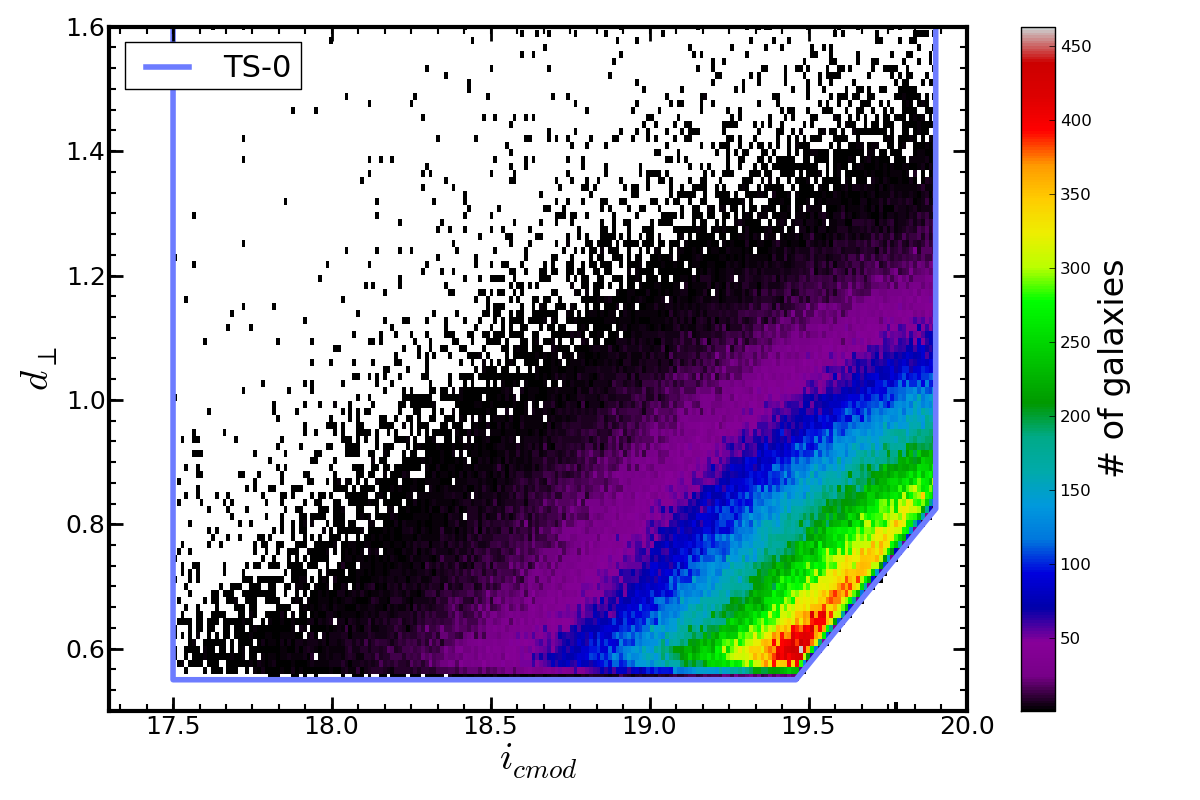}
\caption{The density of galaxies in the CMASS sample in colour-magnitude
 space. The
parameter $d_perp$ is defined in Equation \ref{eq:dperp}. The red colour indicates a high density and black shows low density. The solid blue line represents the CMASS target selection criteria.}
\label{fig:CMASSTS}
\end{figure}

We use data included in data release 12 \cite[DR12;][]{Reid2016,Alam2014} of the Sloan Digital Sky Survey  \cite[SDSS;][]{York2000}. SDSS I, II \citep{Abazajian2009} and III \citep{Eisenstein2011} used a drift-scanning mosaic CCD camera \citep{Gunn1998} to image 14555 square degrees of the sky in five photometric bands \citep{Fukugita1996,Smith2002,Doi2010} to a limiting magnitude of $r <22.5$ using the  2.5-m Sloan Telescope \citep{Gunn2006}  at the Apache Point Observatory in New Mexico. The imaging data were processed through a series of SDSS pipelines \citep{Lupton1999,Pier2003,Padmanabhan2008}. \cite{Aihara2011} reprocessed all of the SDSS imaging data in Data Release 8 (DR8). The Baryon Oscillation Spectroscopic survey \cite[BOSS;][]{Dawson2013} was designed to obtain spectra and redshifts for 1.35 million galaxies covering 10,000 square degrees of sky. These galaxies were selected from the SDSS DR8 imaging. \citep{Blanton2003b} developed a tiling algorithm that is adaptive to the density of targets on the sky and this was used for targeting in BOSS. BOSS used double-armed spectrographs \citet{Smee2013} to obtain the spectra. BOSS resulted in a homogeneous data set with a high redshift completeness of more than 97\% over the full survey footprint. The redshift extraction algorithm used in BOSS is described in \citet{Bolton2012}. \citet{Eisenstein2011} provides a summary and \citet{Dawson2013}  provides a detailed description of the survey design.

We use the CMASS sample of galaxies  \citep{Bolton2012} from data release 12 \citep{Alam2014}. The CMASS sample contains 7,65,433 Luminous Red Galaxies (LRGs) covering 9376 square degrees in the redshift range $0.44<z<0.70$, which correspond to an effective volume of 10.8 Gpc$^{3}$. We used co-added spectra for each galaxy in our analysis \footnote{The co-added version of the spectrum used in our analysis can be downloaded from \url{http://data.sdss3.org/sas/dr12/boss/spectro/redux/v5_7_0/spectra/lite/}. The basic description of the SDSS optical spectra  can be found over \url{http://www.sdss.org/dr12/spectro/spectro_basics}}.

\subsubsection{CMASS Target Selection}
The photometrically identified objects in the SDSS imaging catalog (Data Release 8:DR8\footnote{\url{http://www.sdss3.org/dr8}}) are used as the parent sample for selecting the galaxies to be targeted for spectroscopic observations. The parent catalog covered 7606 ${\rm deg}^2$ in the Northern Galactic Cap (NGC) and 3172 ${\rm deg}^2$ in the Southern Galactic Cap (SGC). The photometric sample contains flux observed in five photometric bands ($u,g,r,i,z$). The target selection for the CMASS sample uses two types of magnitude provided by the SDSS imaging pipeline.  The imaging pipeline fits exponential and deVaucouleurs profiles for each of the five photometric band to provide the fluxes $f_{\rm exp}^{\rm band}$ and $f_{\rm deV}^{\rm band}$ respectively. These fluxes are used to define two different kinds of flux, named ``model'' and ``cmodel'' and given by the following equations.

\begin{equation}
f_{\rm mod,cmod}^{\rm band}=(1-P_{\rm mod,cmod})f_{\rm exp}^{\rm band}+P_{\rm mod,cmod}f_{\rm deV}^{\rm band}.
\end{equation}
Here $P_{\rm mod}$ is a real number between 0 and 1, and $P_{\rm cmod}$ is an integer which can be either 0 or 1. The imaging pipeline fits the observed flux to obtain values of $P_{\rm mod,cmod}$. The main difference between model and cmodel flux is that the model flux results from the use of a linear combination of exponential and  deVaucouleurs profiles, whereas the cmodel flux uses the best-fitting profile. The model and cmodel fluxes are converted to magnitudes as follows:

\begin{equation}
{\rm mag_{band}} = 22.5 -2.5\log( f^{\rm band})-C_{\rm extinction},
\label{eq:mag}
\end{equation} 
where fluxes are in nanomaggies and ${\rm mag_{band}}$ can be any of the five photometric bands $u,g,r,i,z$. The $C_{\rm extinction}$ is the galactic extinction correction for the galaxy using the dust maps of \citet{Schlegel1998}.
 The main criteria used in CMASS target selection are as follows:

\begin{align}
17.5 &< i_{\rm cmod} <19.9  \label{eq:TS1} \\
d_\perp &> 0.55  \label{eq:TS2} \\
i_{\rm cmod} &<1.6(d_\perp-0.8)+19.86 \label{eq:TS3}
\end{align}

The CMASS targets are selected to create a constant
  stellar mass sample. A galaxy evolution model incorporating the
  redshift evolution of band magnitudes is used to determine the
  magnitude cuts that lead to the required sample. Hence the selection
  (using model magnitudes as cuts) is applied without any
  K-correction.  There are several other criteria used for the target
  selection but they affect a very small number of objects and are not
  relevant for our study. The full list of target selection rules is
  provided in \citet{Reid2016}. The quantity $i_{\rm cmod}$ is the
  cmodel magnitude for photometric band $i$. The quantity $d_\perp$ is
  a linear combination of the colour $g-r$ and $r-i$ based on model
  magnitude as follows:

\begin{equation}
\label{eq:dperp}
d_\perp=(r_{\rm mod}-i_{\rm mod})-\frac{1}{8}(g_{\rm mod}-r_{\rm mod}),
\end{equation}
where $g_{\rm mod},r_{\rm mod},i_{\rm mod}$ are the model magnitudes for the photometric bands $g,r$ and $i$ respectively. The Figure \ref{fig:CMASSTS} shows the distribution of galaxies in the final CMASS sample (DR12) in the $i_{\rm cmod}-d_\perp$ plane. The solid line shows the target selection  rule as stated in equation \ref{eq:TS1}, \ref{eq:TS2} and \ref{eq:TS3}.

\subsection{Spectro-Photometry}
\begin{figure}
\includegraphics[width=0.5\textwidth]{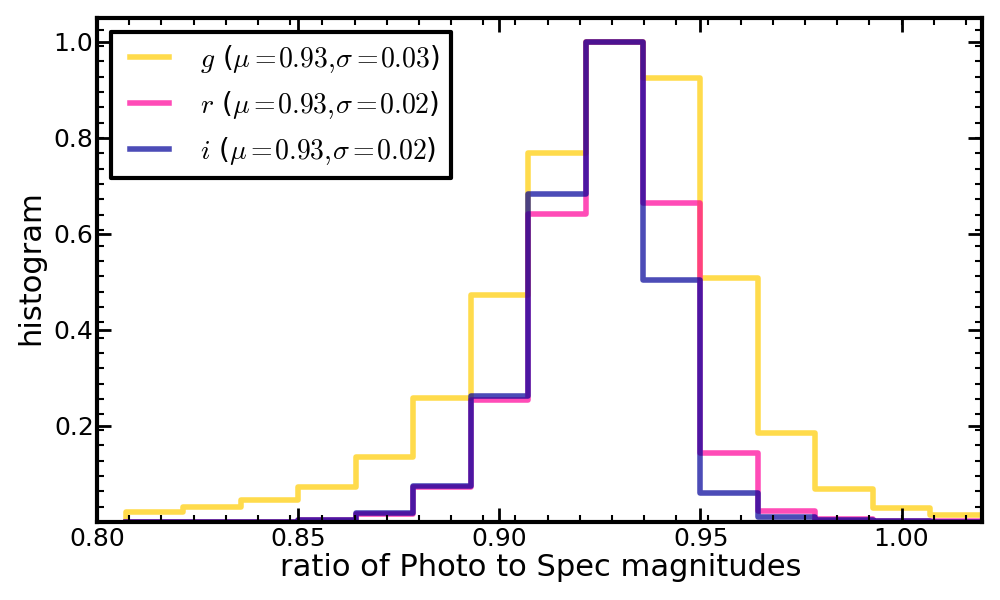}
\caption{The histogram of the ratio of magnitudes from spectra to the photometric magnitude for $g,r$ and $i$ bands. The mean of the ratio is 0.93 which indicates that the magnitudes measured from spectra are larger (flux from spectra is smaller). This is  because the fibres cover only 2\'' which is smaller than the mean size of a galaxy in the sample. This plot also shows that the scatter in this ratio of the two magnitudes is quite small.}
\label{fig:MagR}
\end{figure}

We use SDSS observed SEDs as a template to study the relativistic effects. We assume the observed SEDs are good representation of the galaxy population and treated them as if they were emitted SED of galaxies. We transform each of the observed spectra  according to equation \ref{eq:spectra} for a given $\beta$ and $\theta$. We then obtain the flux in different photometric bands by integrating the spectra with the response function for each band:

\begin{equation}
f_{\rm spec}^{\rm band}=\int d\lambda f(\lambda) R^{\rm band}(\lambda) C^{\rm band},
\end{equation}
where $f(\lambda), R(\lambda)$ represents the flux and photometric band response for wavelength $\lambda$. The parameter $C^{\rm band}$ is the calibration factor which is obtained using the fibre flux of 10,000 galaxies. The calibration factors obtained for $g,r$ and $i$ bands are ($(2.3,3.3,6.1)\time 10^{-3}$ respectively. The fibre flux is another flux provided in the SDSS imaging catalog. It represents the flux obtained in the photometric survey withing the aperture of spectroscopic fibre for each band \footnote{http://www.sdss.org/dr12/algorithms/magnitudes/}. The aperture of $2\,''$ in diameter is assumed for calculating fibre flux, which is appropriate for the BOSS spectrograph.  The spectroscopic flux is converted to magnitude using equation \ref{eq:mag}. The spectroscopic magnitude is typically smaller than the corresponding photometric magnitude because fibres cover only the central part of galaxies. We have found that the spectroscopic magnitudes can be converted to photometric magnitudes using a simple multiplication factor of 0.93. The Figure \ref{fig:MagR} shows the histogram of the ratio of model magnitude to the spectroscopic magnitude. For each of $g,r$ and $i$ band the ratio of magnitudes has mean at 0.93 with a scatter of 0.03 for $g$ band and 0.02 for both $r$ and $i$ band. We therefore obtain the cmodel magnitude from the spectroscopic magnitude using a multiplication factor of 0.93 ($i_{\rm cmod}^{\rm spec}=0.93 i^{\rm spec}$).

\subsection{Magnitude and colour evolution}
\begin{figure}
\includegraphics[width=0.5\textwidth]{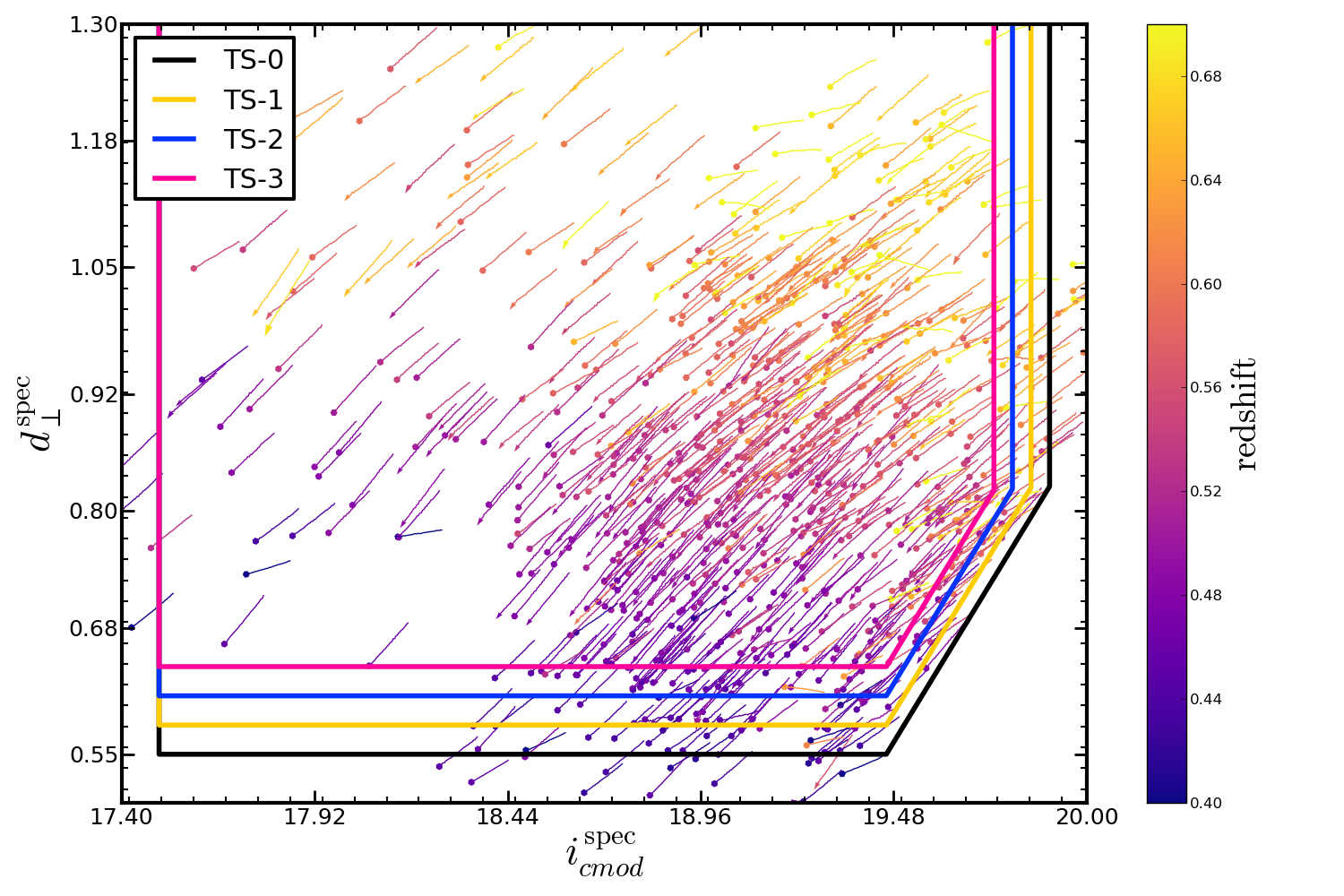}
\caption{The effects of galaxy motion on  observed galaxy colour and magnitude. The solid thick lines of different colours show the different versions of our target selection criteria. The black solid line shows the CMASS original target selection. Other solid lines shows the variant of CMASS target selection described in equation \ref{eq:TSn}. Each line with an arrow head shows how an individual galaxy will move in this space as we assign it a different velocity. The arrow-head shows the observed colour-magnitude when galaxies are moving towards the observer with a speed of 3000 ${\rm km\,s}^{-1}$ and the tail point shows the colour-magnitude when it moves with  speed of 3000 ${\rm km\,s}^{-1}$ away from the observer. The colour of the arrow itself
indicates the redshift of the galaxy. Note that at small redshift a galaxy moving towards observer will cross the colour cut to move out of the sample whereas at higher redshift the galaxy moving towards us with become brighter and cross the lower magnitude cut to move inside the sample.  Note that we only show  a very small illustrative sub-sample of the full CMASS dataset, and we restrict ourselves to velocity directions directly aligned with the line-of-sight. It should be also noted that the effects shown in this plot are exaggerated by roughly an
order of magnitude  compared to the typical case for galaxies, as we are showing results for galaxy velocities as high as 3000 ${\rm km\,s}^{-1}$.}
\label{fig:trace}
\end{figure}

The local gravitational interactions of galaxies causes them to have peculiar velocities. These peculiar velocities cause the observed SEDs of galaxies to be different from the true SEDs. This can change the observed magnitude and colour of galaxies. We systematically investigate these changes for grid of peculiar velocity magnitudes and directions from the line-of-sight. We transform the observed spectra of each galaxy using $\beta$ values between -0.01 and 0.01 and $\theta$ between $0^\circ$ and $90^\circ$. We find that adding relativistic effects to spectra shifts the galaxies in the target selection plane. Not suprisingly, these shifts in colour are sensitive to the galaxy spectra themselves
and therefore depend on the stellar mass and redshift of  galaxies. The Figure \ref{fig:trace} shows the tracks of galaxies in the target selection colour-magnitude  plane. Each line with an arrowhead shows the path followed by the galaxies in the sample as peculiar velocity is varied. The tail of the line corresponds to the colour-magnitude of the galaxy when it is moving away from the observer with $\beta=-0.01$ (speed of 3000 ${\rm km\,s}^{-1}$) and the arrowhead correspond to the case when it is moving towards the observer with the same speed (i.e. we are showing the difference in assigning $\beta$ from $-0.01$ (tail) to $+0.01$ (head). The colour of the track  indicates the redshift of the galaxy. Note that in the plot 
we only show  a very small illustrative sub-sample of the full CMASS dataset,
and we restrict ourselves to velocity directions directly aligned with the line-of-sight. The black thick solid line shows the CMASS target selection as described in equations \ref{eq:TS1}, \ref{eq:TS2} and \ref{eq:TS3}. We also show 3 more restrictive target selection criteria using other solid lines. The target selection criterion TS-n is given by the following equation:

\begin{align}
17.5 &< i_{\rm cmod} <19.9-0.05n \\
d_\perp &> 0.55 +0.03n\\
i_{\rm cmod} &<1.6(d_\perp-0.8-0.05n)+19.86,
\label{eq:TSn}
\end{align}
where $n$ is either 0,1,2, or 3, which represent different target selections TS-0,TS-1,TS-2 and TS-3 respectively. TS-0  is the actual CMASS target selection. Notice that these additional target selections are defined such that the shape of the target selection region in this plane remains unchanged. The tracks of galaxies show that the magnitudes (plotted on the x-axis) decrease (becomes brighter) when galaxies move towards the observer and increase (becomes dimmer) when they moves away  as per our expectation. This leads to galaxies at higher redshifts which are close to the magnitude limit of the target selection being moved inside the sample  when their velocity is towards the observer and being moved outside while their velocity is away. The colour cuts can however reverse this trend as shown by the galaxies close to the lower limit of $d_\perp$, which are at lower redshifts. These galaxies move inside the sample when they have velocities  away from the observer and moves outside the sample with velocities towards the observer. It should be also noted that the effects shown in this plot are exaggerated by roughly an
order of magnitude  compared to the typical case for galaxies, as we are showing results for galaxy velocities as high as 3000 ${\rm km\,s}^{-1}$.

\subsection{Impact on Final Obtained Sample}
\label{sec:wrel}
\begin{figure*}
\includegraphics[width=0.98\textwidth]{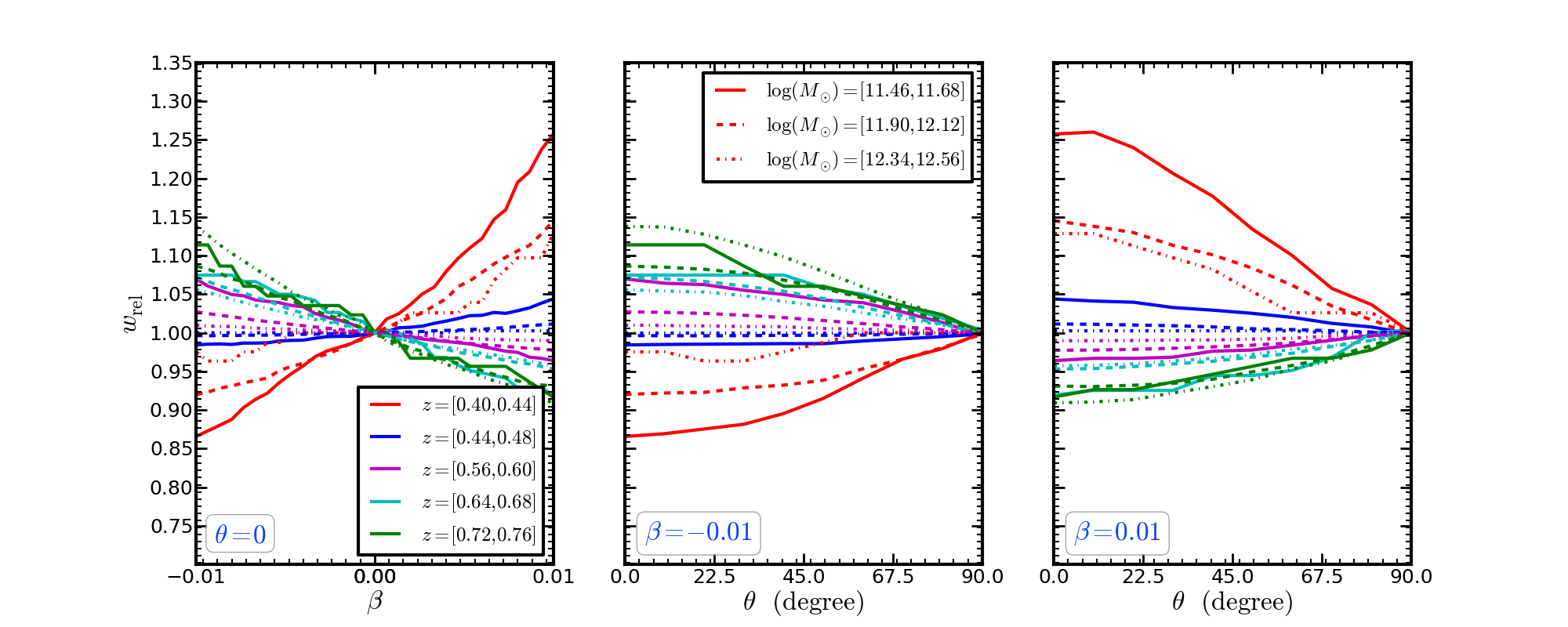}
\caption{The relativistic weights for a galaxy given its redshift, stellar mass and velocity vector. The different colours indicate different redshift bins and different line-styles indicate different stellar mass bins. The left panel shows the $w_{\rm rel}$ with velocity of the galaxy in units  of  the speed of light along line-of-sight. The central and right panel shows the weight dependence on the direction of velocity from line-of-sight for $\beta=-0.01$ (v=3000 ${\rm km\,s}^{-1}$ away from observer) and $\beta=0.01$ (v=3000 ${\rm km\,s}^{-1}$ towards the observer) respectively.}
\label{fig:wrel}
\end{figure*}

Because the peculiar velocities of galaxies vary spatially, the
relativistic effects will spatially modulate the observed SEDs of
galaxies, which will in turn affect the observed magnitudes and
colours. Therefore, a fraction of galaxies with colours and magnitudes
originally within our target selection will move out of the sample and
also some galaxies from outside the sample will move into it. This
affects the observed number density of galaxies in the final
sample. The modulations introduced in the observed number density will
also be correlated with several other properties of galaxies for
example stellar mass, redshift and velocity. In order to quantify
these effects, we bin our sample in redshift and stellar mass. We
create 10 bins in redshift between 0.4 and 0.8 and 10 bins in
logarithm of stellar mass between $10^{10.8} M_{\odot}$ and $10^{13}
M_{\odot}$.  Naively one might think that the sample
  doesn't contain information on galaxies lost due to such effects
  and it should be impossible to correct for these lost galaxies. However,
  this is not the case if we work 
under the assumption that the lost (and extra) galaxies are part
  of the same distribution, and that galaxies properties and dynamics
  are smoothly varying. As long as we are not dominated by noise where
  we overfit small fluctuations in small bins of properties our
  results should be independent of binning used in the sample.  For
each stellar mass and redshift bin, we compute the initial number of
galaxies ($N_{\rm TS}^i$) in the sample. We then transform the
galaxies as if they were moving with velocity $v=\beta c$ along a
direction at angle $\theta$ from the line-of-sight. We then reapply
the target selection boundaries to count the final number of galaxies
in the sample ($N_{\rm TS}^f$). The relativistic effects due to
peculiar motion of galaxies imply that the number of galaxies in the
observed sample will be multiplied by the fraction $N_{\rm
  TS}^f/N_{\rm TS}^i$. Therefore, in clustering analysis if we would
like to compensate for the number density modulation due to
relativistic effects we should weight each galaxy by $w_{\rm rel}$,
where
\begin{equation}
w_{\rm rel}=N_{\rm TS}^i/N_{\rm TS}^f
\end{equation}

We have obtained the $w_{\rm rel}$ for each bin as a function of $\beta$ and $\theta$ of the galaxy. Figure \ref{fig:wrel} shows the weights obtained for some of the redshift and stellar mass bins as the function of $\beta$ and $\theta$. The different colours correspond to different redshift bins, while the different line styles correspond to different stellar mass bins. The left panel shows $w_{\rm rel}$ with $\beta$ between  $-0.01$ and $0.01$ and $\theta=0$. The value $\beta=-0.01$ corresponds to galaxies moving with a speed 3000 ${\rm km\,s}^{-1}$ away from the observer and $\beta=0.01$  galaxies moving at 3000 ${\rm km\,s}^{-1}$ towards the observer. At higher redshifts the galaxies moving towards the observer (positive $\beta$) have weight smaller than 1. They will appear brighter and hence will be seen in larger number than if they were at rest with respect to the observer. The weight in this case is therefore smaller than unity, to compensate for the higher number of observed galaxies. The weights vary with stellar mass, galaxies with higher stellar mass having larger weights. 

These trends change for lower redshifts however. Below approximately $z=0.5$, galaxies moving towards the observer have weights larger than 1. This is due the fact that the galaxies at lower redshift are less likely to be close to  the magnitude limit of the sample than they are to the colour cut. When they move 
towards the observer they cross through the colour cut and out of the sample.
This causes a reverse trend with $\beta$ which is different to that at higher
redshifts. This can be seen in Figure \ref{fig:trace} by following the tracks of these galaxies as $\beta$ is varied. The middle and right panels of Figure \ref{fig:wrel} shows the dependence of $w_{\rm rel}$ on the direction of galaxy velocity $\theta$ for velocities with positive and negative $\beta$. 
These results show the importance of considering the full velocity vector 
rather than just the line-of-sight component.

\subsection{Predicting the galaxy peculiar velocities}
\label{sec:reconvel}

\begin{figure}
\includegraphics[width=0.48\textwidth]{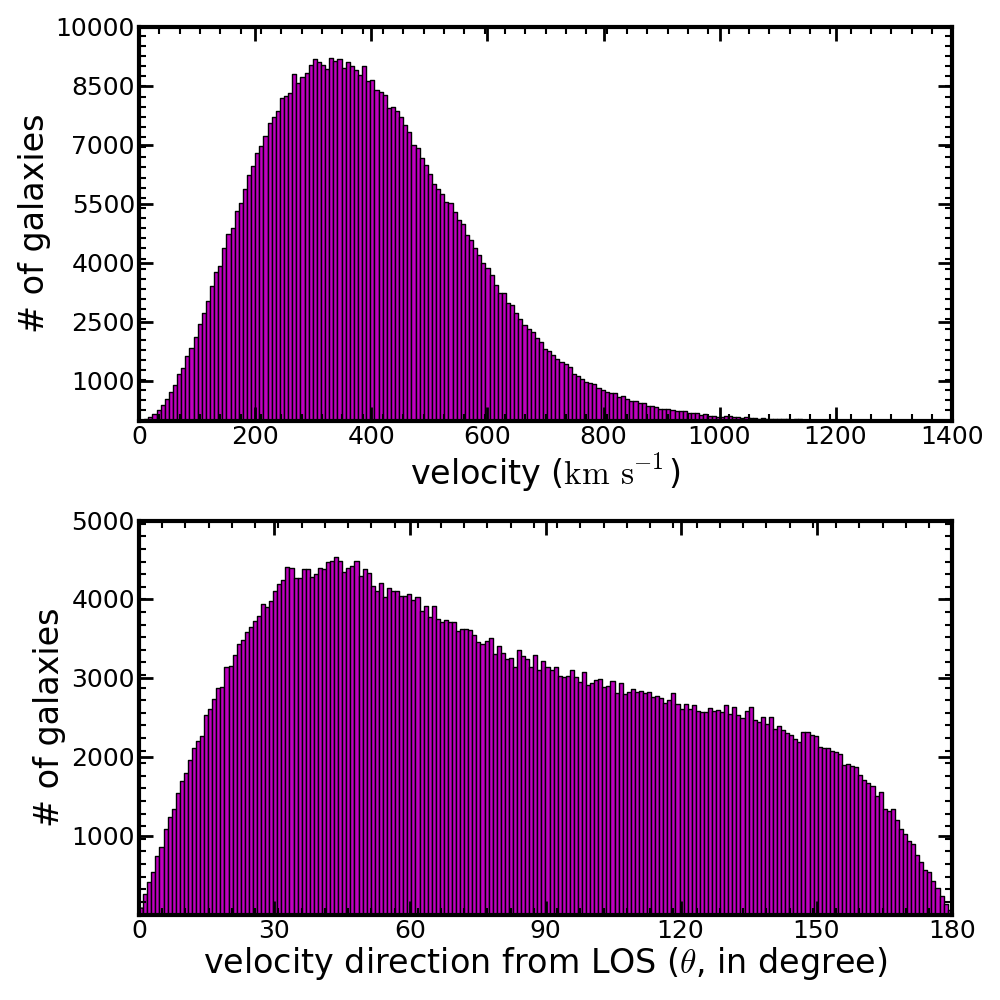}
\caption{Estimated galaxy peculiar velocities in the SDSS III CMASS galaxy redshift sample. The velocity vectors for each galaxy were estimated using a perturbation theory based reconstruction algorithm. The top panel shows the distribution of the magnitudes of galaxy velocities in the sample. The bottom panel shows the distribution of velocity directions, where $\theta=0^\circ$ indicates that a galaxy is moving along line of sight away from observer and  $\theta=180^\circ$  that the  galaxy is moving directly towards the observer.}
\label{fig:vel}
\end{figure}

In order to associate relativistic weights to each individual galaxy, the galaxy velocity is required. We estimate the velocity for each galaxy in the sample using a reconstruction approach. We use a publicly available reconstruction code\footnote{github repo: \url{https://github.com/martinjameswhite/recon_code/}} which estimates the velocities of galaxy in our sample using perturbation theory \citep{White2015code,White2015theory}. The reconstruction code first computes the number density ($\rho$) of galaxies on a grid using a cloud-in-cell assignment scheme . The number density is then converted to density contrast ($\delta$) which is divided by a large scale bias $b$ to yield the mass fluctuation
in the cell. We use the value $b=2.1$ measured in our analysis  \citet[see companion paper:][]{Alam2016Measurement}. This  mass fluctuation is then smoothed using a Gaussian kernel of width $R_f$ (the smoothing scale). Our chosen
value of $R_{f}=10$ h$^{-1}$Mpc is motivated by the results of \citep{Vargas2015}. The reconstruction code then solves for the displacement field
\citep{Zeldovich1970} and provides the displaced position for each galaxy \citep{White2015code}.  We use the displaced position to obtain the peculiar velocities of galaxies using following equation:

\begin{equation}
\vec{v}=afH (\vec{r}_{obs}-\vec{r}_{\rm recon}),
\end{equation} 
where $H=100$ (h$^{-1}$Mpc)/${\rm km\,s}^{-1}$, $a=1/(1+z)$ is the scale factor. We approximate the linear growth rate of perturbations $f=d\ln D/d\ln a$  as $f=\Omega_m(z)^{0.55}$.   Figure \ref{fig:vel} shows the distribution of galaxy velocities obtained using this procedure. In the top panel it can be seen that  most of the galaxies have velocities between $200-600 \, {\rm km\,s}^{-1}$. The bottom panel shows the distribution of the angles between the velocities in the line of sight. The detailed shape of this distribution depends on the geometry the survey.

In an isotropic universe we would expect the velocity
  distribution to be isotropic. This in spherical polar
  coordinates would yield a sin function for the velocity distribution with
  angle. Since the survey geometry is a cone we are
  sub-sampling a cone to estimate the velocity distribution. Simply
  sub-sampling any part of universe with any geometric shape shouldn't
  change the velocity distribution either. However, we
  are estimating the velocity using the sub-sampled galaxy
  distribution. While solving the Poisson equation the effect
of   missing galaxies  will alter the estimated velocities. In the
  simplest picture, since we sample a cone, this implies an area increasing
  as we move farther from the observer, and so more galaxies should be moving
  away from the observer than towards the observer. We believe this to
  be the reason for the sloping distribution  between 30 and 150 degree in the
  lower panel of Figure \ref{fig:vel}. The angle zero degrees is
that pointing
  away from observer and 180 degrees denotes
 pointing towards the observer in our
  convention. 

We note that these velocities are predicted using
perturbation theory which is not accurate on small scales where
non-linear clustering occurs. On scales below our smoothing scale, a
number of galaxies will be moving significantly faster than the
predicted velocity. This will be particularly true in virialised
objects such as galaxy clusters. Our estimate of the strength of
relativistic effects will therefore tend to be an underestimate. We
also note that the fact that the velocities are predicted using
already modulated field will introduce a second order relativistic
correction which we expect to be much smaller and leave for future
studies.

\begin{figure}
\includegraphics[width=0.48\textwidth]{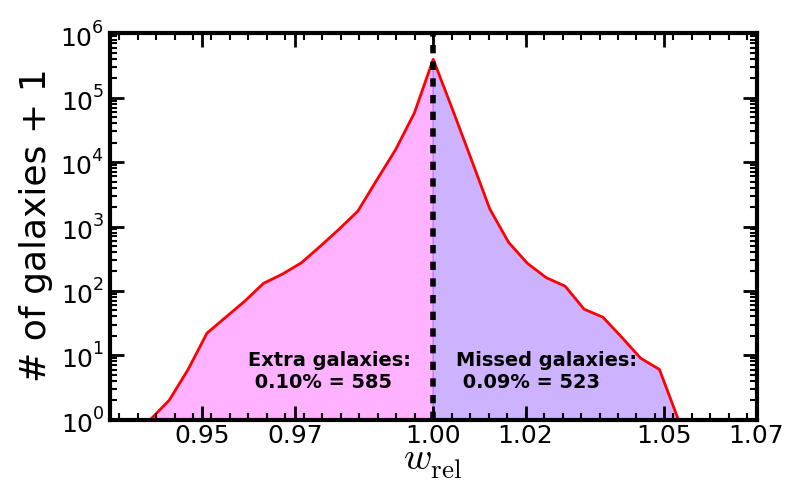}
\caption{The distribution of the relativistic weights
$w_{\rm rel}$ for the CMASS galaxy
redshift sample. The x-axis is $w_{\rm rel}$ and the y-axis displays the binned number of galaxies on logarithmic scale. The galaxies with $w_{\rm rel}<1$ have higher probability of being in the sample. We estimate that 0.16\% more such galaxies have been added to the sample because of their peculiar velocities. Galaxies  with weights $w_{\rm rel}>1$ have a lower probability of being in the sample. From these we calculate that 0.11\% of the sample which would be have been within the colour-magnitude cuts is excluded because of the effect of peculiar velocities.  }
\label{fig:wrel-CMASS}
\end{figure}

\subsection{Impact on Clustering}
\begin{figure}
\includegraphics[width=0.48\textwidth]{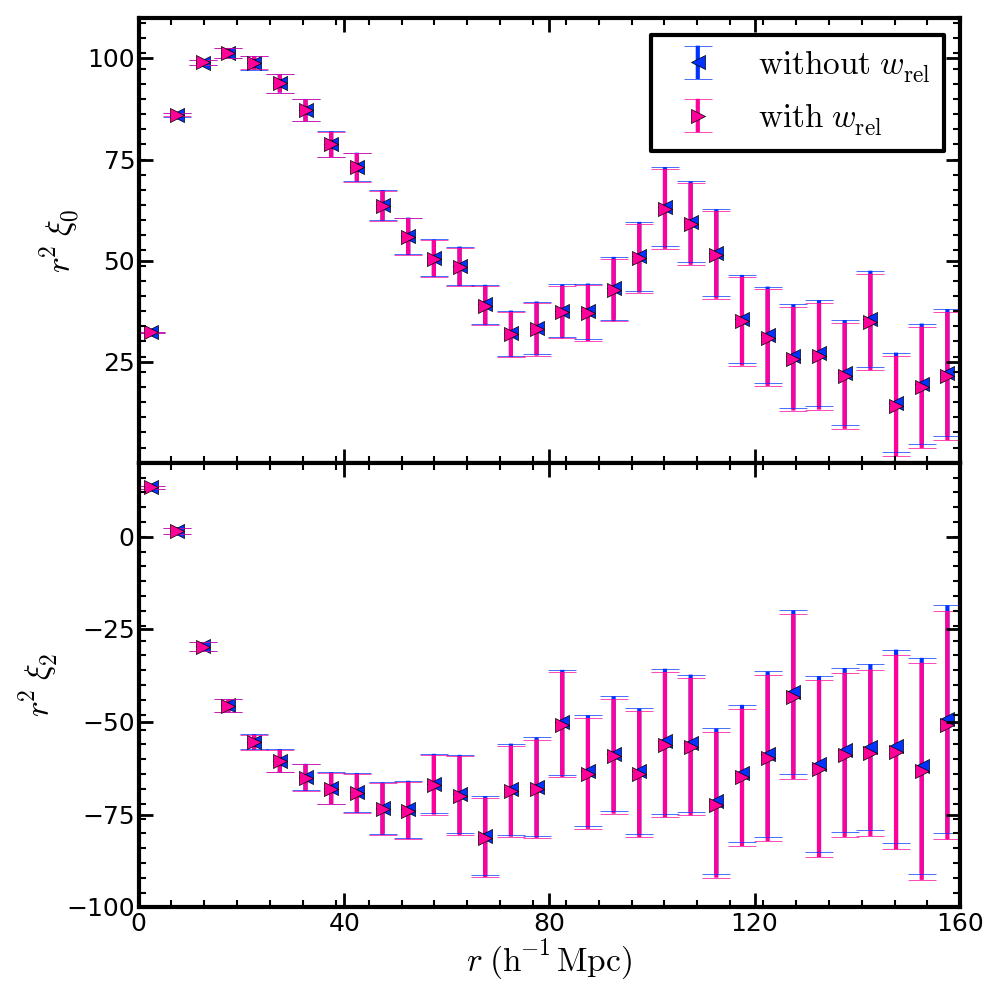}
\caption{The two point galaxy auto-correlation function with and
without the effect of relativistic weights. The top panel shows the monopole and the bottom panel shows the quadruple moment of the correlation function. The blue points represents the measurement without relativistic weight and the magenta points are with the relativistic weight correction. 
}
\label{fig:xi02}
\end{figure}

We now examine how relativistic sample selection effects alter the results of 
standard clustering analyses of the CMASS galaxy redshift sample. 
We use the observational data for the CMASS sample to compute the weights $w_{\rm rel}$ which compensate each galaxy for the effect of Doppler shifting and
beaming (see section \ref{sec:wrel}). These weights are a function of the redshift, stellar mass and velocity vector of the galaxy. The relativistic
correction therefore involves applying the weights before computing the two-point clustering of the galaxy sample. The galaxy catalog contains the redshift and stellar mass of each of the galaxy. We estimate the velocity vector of the galaxy using the perturbation theory approach described in section \ref{sec:reconvel}. Figure \ref{fig:wrel-CMASS} shows the distribution of $w_{\rm rel}$ in the CMASS sample. We can see that the distribution of weights is not symmetric due to the fact that the luminosity function is non-uniform and hence there are more galaxies which scatters into the sample compared to those that scatters out of the sample. We estimate that around 0.10\% ($\sim 585$ galaxies) of the CMASS sample should not have been targeted and around 0.09\% ($\sim 523$ galaxies)  should have been in the sample, but were not observed. 

 We have computed the two-point clustering of CMASS with and without the
 relativistic weights. We use the Landy-szalay \citep{LandySzalay93} estimator, and the results are shown in  Figure \ref{fig:xi02}. The top panel shows the monopole of the correlation function and the bottom panel the quadruple moment.The error bars on the clustering were computed by dividing the entire sample into 61 jackknife regions, see \citet{Alam2016Measurement} for more details. We find that the effects of these weights are much smaller than the statistical errors on the clustering measurement. 

We therefore do not expect that any of the standard large scale structure analyses(such as BAO measurement or redshift space distortions) will show significant effects in current surveys. We should bear in mind though, that as the samples get larger and probe fainter magnitudes these effects might start to become more important for future surveys.

\section{Conclusion}
We have used the SDSS III BOSS CMASS galaxy sample to examine the impact of relativistic effects on observed galaxy SEDs. We have discussed how the effects 
on SEDs will translate to observed fluxes and hence will impact the target 
selection of galaxy redshift surveys. We have found that galaxies can  move both in and out of the sample depending on their peculiar motion. We have investigated  these effects for the CMASS
target selection as a function of redshift, stellar mass, magnitude and direction of galaxy velocity. In order to 
estimate the effect on clustering statistics, we have also
used perturbation theory to predict the galaxy velocities from
the galaxy density field. These velocities provide the information we
need to gauge the impact of relativistic effects on individual galaxies.

We have computed weights that can be used to cancel out the relativistic
effects on target selection.
We studied the galaxy two-point correlation function with and without these weights, finding an impact on the clustering signal which is much smaller than the current statistical errors. This should not therefore affect  current large scale structure analyses such as
 baryon acoustic oscillation measurement or estimates of 
the growth rate from redshift space  distortions. We expect that these effects will be more significant when one is looking at galaxy clustering weighted by one of the properties which are affected by relativistic effects such as luminosity, photometric magnitude etc. We also expect these effects to be more significant when surveys are deeper and hence future surveys should be analyzed with such effects in mind.

One of the main motivations to study these effects is to understand how relativistic beaming and doppler shift modulate the density field and change galaxy
clustering. 
If clustering statistics are chosen carefully and galaxy samples are large
enough, then these effects can in principle be detected.
\citep{Kaiser2013} has shown that these effects can contribute to
the asymmetry in galaxy clustering around clusters which is used to 
infer the gravitional redshift profile  \citep[e.g., ][]{Cappi1995, Kim2004, Wojtak2011, zhao2013, Sadeh2015}. Relativistic
effects on large-scale clustering 
have also been computed using
perturbation theory in full General Relativity
 \citep[e.g., ][]{McDonald2009, Yoo2012, Bonvin2014b}.   
The results in our paper have motivated the form of the beaming effect included in a companion paper \citet{Zhu2016Nbody}. We have applied them to
N-body simulations in order to estimate the line-of-sight asymmetry in the 
non-linear scale cross-correlation function of two galaxy populations with 
different halo masses. 
The models are also used in our other companion paper 
\citet{Alam2016Measurement}, which provides the first measurement of 
line-of-sight asymmetry in the CMASS sample.

\section*{Acknowledgments}
This work was supported by NSF grant AST1412966.  SA is also supported by the European Research Council through the COSFORM Research Grant (\#670193). SA and SH are supported by NASA grants 12-EUCLID11-0004 during part of this study.  We would like to thank Ayesha Fatima for going through the early draft and helping us making the text much more clear. 

SDSS-III is managed by the Astrophysical Research Consortium for the Participating Institutions of the SDSS-III Collaboration including the University of Arizona, the Brazilian Participation Group, Brookhaven National Laboratory, Carnegie Mellon University, University of Florida, the French Participation Group, the German Participation Group, Harvard University, the Instituto de As trofisica de Canarias, the Michigan State/Notre Dame/JINA Participation Group, Johns Hopkins University, Lawrence Berkeley National Laboratory, Max Planck Institute for Astrophysics, Max Planck Institute for Extraterrestrial Physics, New Mexico State University, New York University, Ohio State University, Pennsylvania State University, University of Portsmouth, Princeton University, the Spanish Participation Group, University of Tokyo, University of Utah, Vanderbilt University, University of Virginia, University of Washington, and Yale University.


\bibliography{../Master_Shadab.bib}
\bibliographystyle{mnras}

\label{lastpage}

\end{document}